\documentclass[11pt]{article}
\usepackage{graphicx, bm}
\usepackage{amsmath}
\usepackage{color}

\newcommand{\slv}{\raise.15ex\hbox{$/$}\kern-.53em\hbox{$v$}}
\newcommand{\slF}{\raise.15ex\hbox{$/$}\kern-.53em\hbox{$F$}}
\newcommand{\sll}{\raise.15ex\hbox{$/$}\kern-.40em\hbox{$l$}}
\newcommand{\slP}{\raise.15ex\hbox{$/$}\kern-.53em\hbox{$P$}}
\newcommand{\slp}{\raise.15ex\hbox{$/$}\kern-.53em\hbox{$p$}}
\newcommand{\sln}{\raise.15ex\hbox{$/$}\kern-.53em\hbox{$n$}}
\newcommand{\slq}{\raise.15ex\hbox{$/$}\kern-.53em\hbox{$q$}}
\newcommand{\slR}{\raise.15ex\hbox{$/$}\kern-.53em\hbox{$R$}}
\newcommand{\slz}{\raise.15ex\hbox{$/$}\kern-.53em\hbox{$Z$}}
\newcommand{\slzbar}{\raise.15ex\hbox{$/$}\kern-.53em\hbox{$\bar{Z}$}}
\newcommand{\slQ}{\raise.15ex\hbox{$/$}\kern-.53em\hbox{$Q$}}
\newcommand{\slK}{\raise.15ex\hbox{$/$}\kern-.53em\hbox{$K$}}
\newcommand{\slk}{\raise.15ex\hbox{$/$}\kern-.53em\hbox{$k$}}
\newcommand{\slkone}{\raise.15ex\hbox{$/$}\kern-.53em\hbox{$k_1$}}
\newcommand{\slktwo}{\raise.15ex\hbox{$/$}\kern-.53em\hbox{$k_2$}}
\newcommand{\slkthree}{\raise.15ex\hbox{$/$}\kern-.53em\hbox{$k_3$}}
\newcommand{\slD}{\raise.15ex\hbox{$/$}\kern-.53em\hbox{$D$}}
\newcommand{\slC}{\raise.15ex\hbox{$/$}\kern-.53em\hbox{$C$}}
\newcommand{\slA}{\raise.15ex\hbox{$/$}\kern-.53em\hbox{$A$}}
\newcommand{\slSigma}{\raise.15ex\hbox{$/$}\kern-.53em\hbox{$\Sigma$}}
\newcommand{\slpartial}{\raise.15ex\hbox{$/$}\kern-.53em\hbox{$\partial$}}
\newcommand{\slcalP}{\raise.15ex\hbox{$/$}\kern-.63em\hbox{$\cal P$}}

\newcommand{\bx}{{\bm x}}

\newcommand{\tr}{\text{tr}}

\def\be{\begin{equation}}
\def\ee{\end{equation}}
\def\bea{\begin{eqnarray}}
\def\eea{\end{eqnarray}}

\begin{document}

\title{ \bf \Large Polarized 3 parton production in inclusive DIS at small $x$}
\author{
Alejandro~Ayala$^{1,2}$, Martin~Hentschinski$^{1,3}$,
  Jamal~Jalilian-Marian$^{4,5}$, \\   Maria~Elena~Tejeda-Yeomans$^6$
 \bigskip \\  
{ \normalsize $^1$Instituto de Ciencias Nucleares, Universidad Nacional Aut\'onoma
  de M\'exico,} 
 \\
{\normalsize Apartado Postal 70-543, Ciudad de M\'exico 04510, Mexico } \\
{\normalsize $^2$Centre for Theoretical and Mathematical Physics, and Department 
of Physics}, \\ 
{\normalsize University of Cape Town, Rondebosch 7700, South Africa}\\
{\normalsize $^3$Facultad de Ciencias F\'isico Matem\'aticas, Benem\'erita
  Universidad Aut\'onoma de Puebla, } \\ 
{\normalsize  Puebla 1152, Mexico}\\
{\normalsize $^4$Department of Natural Sciences, Baruch College, CUNY,
}
 \\ {\normalsize
17 Lexington Avenue, New York, NY 10010, USA}\\
{\normalsize $^5$CUNY Graduate Center, 365 Fifth Avenue, New York, NY 10016, USA}\\
{ \normalsize $^6$Departamento de F\1sica, Universidad de Sonora, Boulevard
Luis Encinas J. y Rosales, } \\ 
{\normalsize Colonia Centro, Hermosillo, Sonora 83000,
Mexico}
}

\date{}
\maketitle

\begin{abstract}
  \noindent Azimuthal angular correlations between produced
  hadrons/jets in high energy collisions are a sensitive probe of the
  dynamics of QCD at small $x$. Here we derive the triple differential
  cross section for inclusive production of $3$ polarized partons in
  DIS at small $x$. The target proton or nucleus is described using
  the Color Glass Condensate (CGC) formalism. The resulting
  expressions are used to study azimuthal angular correlations between
  produced partons in order to probe the gluon structure of the target
  hadron or nucleus. Our analytic expressions can also be used to
  calculate the real part of the Next to Leading Order (NLO)
  corrections to di-hadron production in DIS by integrating out one of
  the three final state partons.
\end{abstract}

Gluon saturation in QCD at small Bjorken $x$ was proposed by Gribov,
Levin and Ryskin as a dynamical mechanism by which perturbative
unitarity of QCD cross sections at high energy (small $x$) is
restored~\cite{glr}. McLerran and Venugopalan~\cite{mv} formulated an
effective action approach to gluon saturation which describes a high
energy hadron or nucleus as a Color Glass Condensate (CGC); a state of
high occupancy number which is a weakly-coupled yet non-perturbative
system of gluons characterized by a semi-hard scale $Q_s$, called the
saturation scale. This is a systematic approach which allows to apply
semi-classical methods to particle production in high energy hadronic
collisions. While there is a large body of evidence pointing to the
importance of saturation physics in the small $x$ regime of high
energy collisions ~\cite{cgc-reviews}, more studies are needed to
constrain the parameters of the approach and to clarify further its
kinematic limits.\\

The bulk of current studies of Deep Inelastic Scattering (DIS) of
electrons on a target proton or nucleus \cite{cgc-dis, Beuf:2011xd}
concentrates on inclusive observables, such as (diffractive) structure
functions $F_{2}$ and $F_L$, which provide access to 2-point
correlations (of Wilson lines) in the target as encoded in the
extracted color dipole factors. Dynamics of gluon saturation is
however far richer than that contained in 2-point functions:
observables in the CGC approach are given in terms of multi-point
correlators of Wilson lines, the most prominent one being the
Quadrupole -- the 4-point correlation function of Wilson lines which
appears in multi-parton production processes in DIS and pA collisions
\cite{cgc-double}. These higher point correlators carry 
much more information about the dynamics of gluon saturation than
dipoles.  However, unlike the dipoles, higher point correlators are
experimentally not well-constrained. To access them experimentally, it
is needed to study observables with multi-particle final states in a
clean experimental environment such as DIS. A first example is
azimuthal angular correlation of two hadrons \cite{az-ang} which
provides access to the 4-point correlation function and constitutes a
key process in saturation searches at future Electron Ion Colliders,
see {\it e.g.} \cite{zalx}.\\

In this Letter we propose to use azimuthal angular correlations of $3$
partons in inclusive DIS to explore the dynamics of saturated partonic
matter. The novel feature of this process is that, since it involves
two relative angles between the three produced partons, it offers an
additional handle compared to di-hadron azimuthal angular
correlations, which involve only one relative angle.  Moreover, unlike
$2$ parton production, the $3$ parton cross-section depends
non-linearly on both quadrupoles and dipoles and allows therefore to
access different aspects of those correlators; it therefore acts also
as a complementary tool for their exploration.\\

In the following we present our results for the triple differential
cross section for inclusive production of three partons in DIS. Our
expression keeps the full helicity dependence of initial and final
state particles and therefore can be used in principle to study
fragmentation of polarized partons into hadrons \cite{collins}. In
order to exhibit the effect of gluon saturation on azimuthal angular
correlations, we perform a first study of the angular structure of the
three produced partons for a particular angular configuration, in the
transverse momentum region just above the saturation scale. This
allows us to work in the dilute limit of the CGC, but still exhibits
the main features of the saturation dynamics.  We find that saturation
effects reduce the magnitude of the correlation peak while
simultaneously widening it.  We further stress that our analytic
expressions can be used to compute the real contributions to the Next
to Leading Order (NLO) corrections~\cite{rcBK} to inclusive di-hadron
production in
DIS~\cite{zalx}.\\

\begin{figure*}
{\centering {
\includegraphics[scale=0.21]{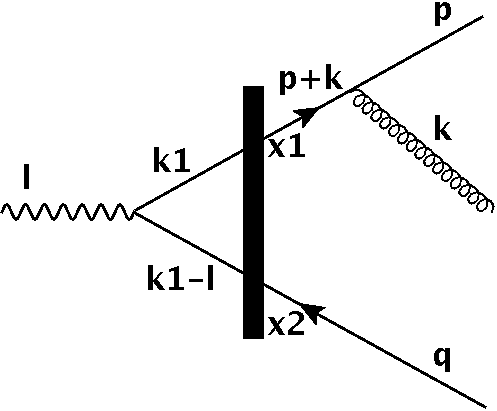}\hspace{.1cm}
\includegraphics[scale=0.21]{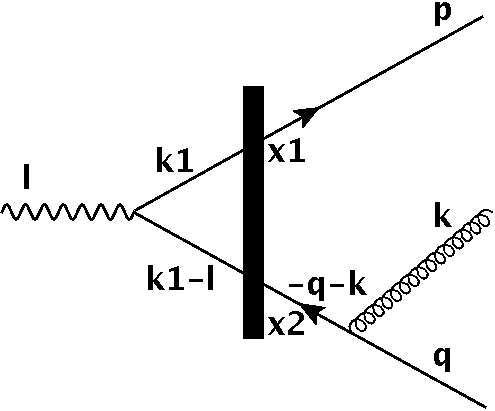}\hspace{.1cm}
\includegraphics[scale=0.21]{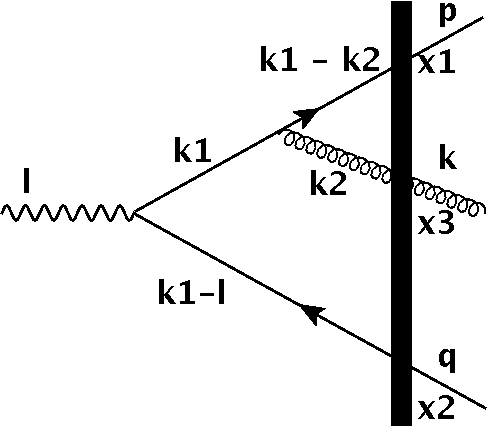}\hspace{.1cm}
\includegraphics[scale=0.21]{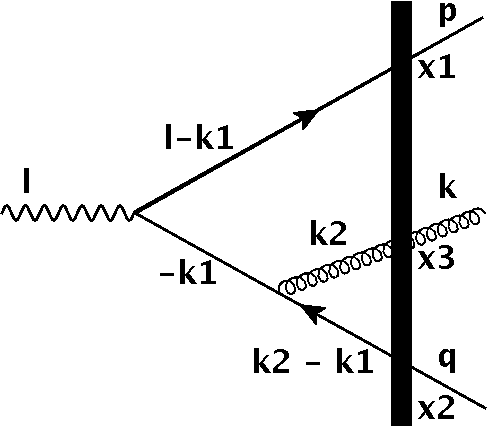} 
}}
\caption{\small  $3$-parton production diagrams.  The solid thick line represents interactions with the target (shock wave). The arrows indicate the direction of fermion charge flow. The photon momentum is incoming whereas all the final state momenta are outgoing.
} 
\label{fig1}
\end{figure*}

We consider the process depicted in Fig.~\ref{fig1} 
\begin{align}
  \label{eq:process}
  \gamma^*(l) + \text{target}\, (P) & \to q(p) + \bar{q}(q) + g(k) + X,
\end{align}
with photon virtuality $l^2 = -Q^2$. The target is a high energy
hadron/nucleus, represented by a strong background color field (shock
wave) $A^\mu \sim 1/g$ in light-cone gauge $A \cdot n = 0$ with
$A^+(x^-, x_t) = \delta(x^-) \alpha(x_t)$.
Light-cone vectors $n, \bar{n}$ are defined through the four momenta
of virtual photon and target, {\it i.e.} $n \sim P$ and
$n \cdot \bar{n} = 1$ with $v^- \equiv n\cdot v$ for a generic four
vector $v$.  Amplitudes are written in terms of momentum space quark
and gluon propagators in the presence of the background field
\begin{align}
S_{F, il} (p,q) &\equiv 
S_{F, ij}^{(0)}(p)  \, \tau_{F, jk} (p,q)  \, S_{kl}^{(0)} (q), &  
G_{\mu\nu}^{ad} (p,q) &\equiv 
G^{(0),ab}_{\;\;\mu\lambda} (p) \,  \tau_g^{bc} (p,q)  \,   G^{(0), cd, \lambda}_{\;\;\;\; \nu} (q)
\end{align}
with  the conventional free fermion and gluon propagator,
$S_{F, ij}^{(0)} (p) =i \delta_{ij} { / (\slp + i\epsilon)}$ and
$G^{(0), ab}_{\;\;\mu\nu}= i\delta^{ab} { d_{\mu\nu} (k)
  /(k^2+i\epsilon)} $
respectively, and
$d_{\mu\nu} (k) = \left[- g_{\mu\nu} + {(k_\mu n_\nu + k_\nu n_\mu)/
    n\cdot k}\right] $
the light-cone gauge polarization tensor.  Furthermore
\begin{align}
  \label{eq:14}
\tau_{F, ij} (p,q) &\equiv (2\pi) \delta (p^- - q^-) \sln \int \!\!d^2
                     x_t \, \, e^{i (p_t - q_t)\cdot x_t}
 \left[\theta (p^-) V_{ij}(x_t) - \theta (-p^-) V_{ij}^\dagger
  (x_t)\right]\, ,
\notag \\
\tau_g^{ab} (p,q) &\equiv - 4\pi p^-  \delta (p^- - q^-) \int\!\! d^2 x_t\,\, e^{i (p_t - q_t)\cdot x_t}
 \left[\theta (p^-) U^{ab}(x_t) - \theta (-p^-) (U^\dagger)^{ab} (x_t)\right]]\, ,
  \end{align}
where  $U^{ba}(z_t) = 2\tr(t^a V(z_t) t^b V^\dagger(z_t))$ and
\begin{align}
  \label{eq:1}
 V (x_t) \equiv \hat{P} \exp \; \{i g \int_{-\infty}^\infty  dx^- A^+_a (x^-, x_t) \, t_a\}, 
\end{align}
the fundamental SU$(N_c)$ Wilson line.  The four matrix elements
corresponding to Fig.~\ref{fig1} read
\begin{align}
  i{\cal A}_1 = &  (i e) (i g)  \int {d^4 k_1 \over (2\pi)^4} \bar{u}(p)   \gamma^\mu\, t^a\, 
S_F (p+k, k_1)  \gamma^\nu 
  S_F (k_1-l, -q)
 \big[S_F^{(0)} (-q)\big]^{-1}\, v(q) \, \epsilon_\nu (l) \,
   \epsilon^*_\mu (k)\, , 
 \notag \\
i{\cal A}_2 =&  (i e) (i g) \int  {d^4 k_1 \over (2\pi)^4} \bar{u} (p) \, \big[S_F^{(0)} (p)\big]^{-1}S_F (p, k_1)\, 
  \gamma^\nu S_F (k_1-l,-q-k) \, \gamma^\mu\, t^a\,  v(q) \,,
  \epsilon_\nu (l) \, \epsilon^*_\mu (k)\, \notag \\
i{\cal A}_{3}   =&   (i e) (i g) \int {d^4 k_1 \over (2\pi)^4} \int
                 {d^4 k_2 \over (2\pi)^4} \bar{u} (p) \big[S_F^{(0)}
                 (p)\big]^{-1} 
  S_F (p, k_1-k_2)
\gamma^\lambda\,t^c\, S_F^{(0)} (k_1)
 \gamma^\nu 
\notag \\
&   S_F (k_1-l,-q)  \big[S_F^{(0)} (-q)\big]^{-1}\!\!  v(q)  
 \left[G_{\lambda}^{\;\,\delta}\right]^{ca}\!\! (k_2,k)
 \big[G_{\delta}^{(0), \mu} (k) \big]^{-1}   \epsilon_\nu (l)  \epsilon_\mu^* (k)\,  , \notag \\
i{\cal A}_{4} =&  (i e) (i g)
 \int {d^4 k_1 \over (2\pi)^4} \int {d^4 k_2 \over (2\pi)^4} 
 \bar{u} (p) \big[S_F^{(0)} (p)\big]^{-1} S_F (p, l-k_1)\, 
\gamma^\nu\, S_F^{(0)} (-k_1) \gamma^\lambda\, t^c  
\notag \\
&
S_F (k_2-k_1,-q)
 \big[S_F^{(0)} (-q)\big]^{-1}\! \left[G^{\;\,\delta}_\lambda\right]^{ca}\! (k_2,k) 
  \big[G^{(0) \mu}_\delta (k) \big]^{-1} 
\epsilon_\nu (l)  \, 
 \epsilon^*_\mu (k),
\label{eq:calA1234}
\end{align}
where $\epsilon_\nu (l), \epsilon^*_\mu (k)$ denote polarization
vectors of the incoming virtual photon and the outgoing gluon
respectively.  With flux factor $\mathcal{F} = 2 l^-$, photon momentum
fractions $\{z_1, z_2, z_3 \} = \{p^-/l^-, q^-/l^-, k^-/l^-\}$ and the
three-particle phase space
\begin{align}
  \label{eq:10}
   d\Phi^{(3)}& =\frac{1}{(2 \pi)^8} \,  d^4p  \, d^4 q  \,  d^4 k  \, 
                  \delta(p^2)\delta(q^2)\delta(k^2)
\delta(l^-  -  p^-  -
                  k^- - q^-) \notag \\
&=
2l^-  \frac{ d^2 p_t \,  d^2 q_t\,  d^2 k_t }{(64 \pi^4 l^-)^2}
  \frac{dz_1  \,   dz_2 \,  dz_3}{z_1 \,
  z_2 \,  z_3 } \delta(1-\sum_{i=1}^3z_i)
\end{align}
 the differential 3-parton production  cross-section reads
\begin{align}
  \label{eq:2}
  d \sigma & = \frac{1}{\mathcal{F}} \left\langle   |\mathcal{A}(A^+) -  \mathcal{A}(0)|^2  \right\rangle_{A^+} d \Phi^{(3)},
\end{align}
where $\langle \ldots \rangle_{A_+}$ denotes the average over
background field configurations.  The differential cross-section can
be expressed in terms of 6 leading $N_c$ terms and one $N_c$
suppressed term. With dipole and quadrupole given by
 \begin{align}
   \label{eq:17}
   S^{(2)}_{({\bm x}_1 {\bm x}_2)} 
 & \equiv 
\frac{1}{N_c} \tr \left[ V({\bm x}_1) V^{\dagger}( {\bm x}_2) \right],  
&  S^{(4)}_{({\bm x}_1 {\bm x}_2 {\bm x}_3 {\bm x}_4)}
 &\equiv \frac{1}{N_c} \tr \left[ V({\bm x}_1) V^{\dagger}( {\bm x}_2) V({\bm x}_3) V^{\dagger}( {\bm x}_4) \right],
 \end{align}
we find the following set of operators,
\begin{align}
  \label{eq:18}
   N^{(4)}({\bm x}_1,{\bm x}_2,{\bm x}_3,{\bm x}_4) & \equiv  
  1 + S^{(4)}_{({\bm x}_1{\bm x}_2{\bm x}_3{\bm x}_4)}    -  S^{(2)}_{({\bm x}_1{\bm x}_2) }
-  S^{(2)}_{({\bm x}_3{\bm x}_4)} \, ,
\notag \\
 N^{(22)} ({\bm x}_1,{\bm x}_2 |{\bm x}_3,{\bm x}_4)  & \equiv  
 \left[S^{(2)}_{({\bm x}_1 {\bm x}_2)} -1 \right] \left[S^{(2)}_{({\bm x}_3 {\bm x}_4)} -1 \right]
\notag \\
 N^{(24)}({\bm x}_1,{\bm x}_2|{\bm x}_3,{\bm x}_4,{\bm x}_5,{\bm x}_6) 
 & \equiv 
        1  + 
S^{(2)}_{({\bm x}_1 {\bm x}_2)}  S^{(4)}_{({\bm x}_3{\bm x}_4{\bm x}_5{\bm x}_6)}  
   -   S^{(2)}_{({\bm x}_1 {\bm x}_2)}   S^{(2)}_{({\bm x}_3 {\bm x}_6 )}
- 
S^{(2)}_{({\bm x}_4 {\bm x}_5)}
\, ,
\notag \\
   N^{(44)}({\bm x}_1,{\bm x}_2,{\bm x}_3,{\bm x}_4| {\bm x}_5,{\bm x}_6, {\bm x}_7, {\bm x}_8)   & \equiv  
1  + 
 S^{(4)}_{({\bm x}_1{\bm x}_2{\bm x}_3{\bm x}_4) }  S^{(4)}_{({\bm x}_5{\bm x}_6{\bm x}_7{\bm x}_8)   }
 - S^{(2)}_{({\bm x}_1{\bm x}_4)} S^{(2)}_{({\bm x}_5 {\bm x}_8)}     
\notag \\
& \hspace{6cm}
  - S^{(2)}_{({\bm x}_2 {\bm x}_3) }S^{(2)}_{({\bm x}_6 {\bm x}_7) }
.
\end{align}
For diffractive reactions, corresponding to color singlet exchange
between the $q, \bar{q}, g$ state and target, all of the above
quadrupoles are found to factorize into the product of two dipoles,
$ S^{(4)}_{({\bm x}_i{\bm x}_j{\bm x}_k{\bm x}_l) } \to S^{(2)}_{({\bm
    x}_i{\bm x}_l)}S^{(2)}_{({\bm x}_j{\bm x}_k)}$.
With $\alpha_{em}$ and $\alpha_s$ the electromagnetic and strong
coupling constants, and $e_f$ the electro-magnetic charge of the quark
with flavor $f$ we obtain the following leading $N_c$ result:
\begin{align}
  \label{eq:19}
&   \frac{d \sigma^{T,L}}{d^2{\bm p}\, d^2 {\bm k} \, d^2 {\bm q} \, dz_1 dz_2}    = \notag \\
& \quad  =\frac{\alpha_s   \alpha_{em} e_f^2 N_c^2 }{z_1 z_2 z_3 (2 \pi)^2} 
  \prod_{i=1}^3 \prod_{j=1}^3   \int \frac{d^2 {\bm x}_i}{(2 \pi)^2}
 \int   \frac{d^2 {\bm x}_j'}{(2 \pi)^2}   e^{i {\bm p}({\bx_1 - \bx_1'}) + i {\bm q}({\bx_2 - \bx_2'}) + i {\bm k}({\bx_3 - \bx_3'})}
  \notag 
\\ & \qquad  \bigg\langle
(2 \pi)^4\bigg[ \bigg( 
 \delta^{(2)}(\bx_{13})\delta^{(2)}(\bx_{1'3'})\sum_{h, g}  \psi^{T,L}_{1;h,g}({\bm x}_{12})   \psi^{T,L,*}_{1';h,g}({\bm x}_{1'2'}) 
+
\{ 1, 1' \}  \leftrightarrow \{ 2, 2'\}
\bigg)  \notag \\
& 
 \quad  \qquad \cdot   N^{(4)}({\bm x}_1,{\bm x}_1',{\bm x}_2',{\bm x}_2)
+
\bigg( 
 \delta^{(2)}(\bx_{23})\delta^{(2)}(\bx_{1'3'})\sum_{h, g}  \psi^{T,L}_{2;h,g}({\bm x}_{12})   \psi^{T,L,*}_{1';h,g}({\bm x}_{1'2'}) 
\notag \\
& \hspace{6.6cm}
+
\{ 1, 1' \}  \leftrightarrow \{ 2, 2'\}
\bigg)\cdot
 N^{(22)}({\bm x}_1,{\bm x}_1'|{\bm x}_2',{\bm x}_2)
 \bigg]\notag \\
& \quad 
 + (2 \pi)^2\bigg[
\delta^{(2)}(\bx_{13}) \sum_{h, g}  \psi^{T,L}_{1;h,g}({\bm x}_{12})   \psi^{T,L,*}_{3';h,g}({\bm x}_{1'3'}, \bx_{2'3'}) N^{(24)}(\bx_{3'}, \bx_{1'}| \bx_{2'} ,\bx_2 ,\bx_1 ,\bx_{3'}) 
 \notag \\
&  \quad \qquad   +
\{ 1 \}  \leftrightarrow \{ 2\}
+
 \delta^{(2)}(\bx_{1'3'}) \sum_{h, g}  \psi^{T,L}_{3;h,g}({\bm x}_{13}, \bx_{23})   \psi^{T,L,*}_{1';h,g}({\bm x}_{1'2'})
\notag \\
& \hspace{6.4cm} \cdot 
 N^{(24)}(\bx_{1}, \bx_{3}| \bx_{2'} ,\bx_2 ,\bx_3 ,\bx_{1'}) 
 +
\{ 1' \}  \leftrightarrow \{ 2'\}
\bigg]
\notag \\
& \quad 
+ \sum_{h, g}  \psi^{T,L}_{3;h,g}({\bm x}_{13}, \bx_{23})   \psi^{T,L,*}_{3';h,g}({\bm x}_{1'3'}, \bx_{2'3'}) \cdot N^{(44)}(\bx_{1}, \bx_{1'}, \bx_{3'}, \bx_3| \bx_{3} ,\bx_{3'} ,\bx_{2'} ,\bx_{2}) 
 \bigg\rangle_{A^+} \, ,
\end{align}  
where $\psi_{i'} \equiv \psi_{i}$, $i=1, \ldots, 3$ and
$z_3 = 1-z_1 - z_2$. To obtain sub-leading terms in $N_c$ all
operators $N^{(4)}, N^{(22)}, N^{(24)}, N^{(44)}$ are to be replaced
by $1/N_c \cdot N^{(4)} (\bx_1, \bx_{1'}, \bx_{2'}, \bx_2)$. The
super-scripts $T=\pm$ and $L$ refer to transverse and longitudinal
polarizations of the virtual photon while $h,g=\pm$ denote quark and
gluon helicity respectively (due to helicity conservation in mass-less
QCD the helicity of the anti-quark is always opposite to the quark
helicity).  To determine the wave functions $\psi_{i}$,
$i=1, \ldots, 3$ we factorize the color and Wilson-line structures
from the amplitudes in Eq.~\eqref{eq:calA1234} and evaluate Dirac and
Lorentz structures using spinor helicity techniques \cite{dixon},
which provide a powerful alternative for the evaluation of scattering
amplitude in the high energy limit; for details we refer to the paper
in preparation \cite{long:2016}.  With $ \phi_{{ij}}$ the azimuthal
angle of ${\bx}_{ij}$, $i,j=1\ldots, 3$ and
\begin{align}
  \label{eq:16}
X_j^2 & = \bx_{12}^2 (z_j + z_3)
   \left(1-z_j - z_3\right), \quad j=1,2,  & 
  X_3^2 & = z_1 z_2 \bx_{12}^2+  z_1 z_3 \bx_{13}^2+ z_2 z_3 \bx_{23}^2 \, ,
\end{align} we obtain
  \begin{align}
  \label{eq:M_L}
  \psi_{j,hg}^{L}  &= \sqrt{2}  Q  
 K_0\left(Q X_j\right) \cdot a_{j,hg}^{(L)},   &   j &=1,2 
\notag \\
  \psi_{j,hg}^{T}  &= 
\frac{   K_1\left(Q X_j \right) }{-i|\bx_{12}| e^{\mp i \phi _{\bx_{12}}}}
\cdot   a_{j,hg}^{\pm}  &   j &=1,2  \notag \\
 \psi_{3,hg}^{L}  &=
4  \pi i Q  \sqrt{2 z_1 z_2}    K_0\left(Q  X_3 \right)  (a_{3,hg}^{(L)} + a_{4,hg}^{(L)}), \notag \\ 
 \psi_{3,hg}^{T}  &=
4  \pi Q  \sqrt{ z_1 z_2}    \frac{K_1\left(Q  X_3 \right)}{X_3}  (a_{3,hg}^{\pm} + a_{4,hg}^{\pm})\, ,
\end{align}
where the helicity amplitude $a_{k,hg}^{(T,L)}$ is directly extracted
from the amplitude $i\mathcal{A}_k$, $k=1, \ldots 4$. These helicity
amplitudes satisfy the following relations,
\begin{align}
  \label{eq:3}
 & a_{k+1,hg}^{T,L}  =  -  a_{k,-hg}^{T,L} (\{p, \bx_1\} \leftrightarrow \{q, \bx_2\}),\qquad  k = 1,3\, , &
  a_{j,hg}^{T,L}  &= a_{j,-h-g}^{(-T,L)*}, \quad   j=1,\ldots ,4.
\end{align}
which allows to simplify the calculation using a minimal set of helicity amplitudes
\begin{align}
 a^{(L)}_{1,++}&= 
- \frac{(z_1 z_2)^{3/2}  \left(z_1+z_3\right)}{z_3 e^{-i \theta _p} |{\bm p}|-z_1
   e^{-i \theta _k} |{\bm k}|},
 &  a^{(L)}_{1,-+} & = 
-\frac{\sqrt{z_1} z_2^{3/2} \left(z_1+z_3\right){}^2}{z_3 e^{-i \theta _p} |{\bm p}|-z_1
   e^{-i \theta _k} |{\bm k}|},
\notag \\
 a^{(L)}_{3,++} &= \frac{z_1z_2}{|{\bm x}_{13}|e^{-i \phi_{\bx_{13}}}}, &
a^{(L)}_{3,-+}& = \frac{z_2(1-z_2)}{|{\bm x}_{13}|e^{-i \phi_{\bx_{13}}}},
\notag \\
   a^{(+)}_{1,++}&= -\frac{\sqrt{2} (z_1 z_2)^{3/2}}{z_3 e^{-i \theta _p} |{\bm p}|-z_1
   e^{-i \theta _k} |{\bm k}|}, 
&
a^{(+)}_{1,-+} & = \frac{\sqrt{z_1z_2} (z_1 + z_3)^2}{z_3 e^{-i \theta _p} |{\bm p}|-z_1
   e^{-i \theta _k} |{\bm k}|} , \notag \\
 a^{(+)}_{1,+-}& =- \frac{\sqrt{z_1}z_2^{3/2} (z_1 + z_3)}{z_3 e^{i \theta _p} |{\bm p}|-z_1 
   e^{i \theta _k} |{\bm k}|},
&
 a^{(+)}_{1,--}&=- \frac{z_1^{3/2}\sqrt{z_2} (z_1 + z_3)}{z_3 e^{i \theta _p} |{\bm p}|-z_1
   e^{i \theta _k} |{\bm k}|},
\notag \\
a^{(+)}_{3,++} & = \frac{z_1 z_2 (z_3 |\bx_{23}| e^{- i \phi_{\bx_{23}}} - z_1 |\bx_{12}| e^{- i \phi_{\bx_{12}}} )}{(z_1 + z_3)|\bx_{13}| e^{- i \phi_{\bx_{13}}}},
\notag \\
a^{(+)}_{3,+-} & = 
\frac{ z_2 (z_3 |\bx_{23}| e^{- i \phi_{\bx_{23}}} - z_1 |\bx_{12}| e^{- i \phi_{\bx_{12}}} )}{|\bx_{13}| e^{ i \phi_{\bx_{13}}}},
\notag \\
a^{(+)}_{3,-+} &=  \frac{(z_1 + z_3) (z_3 |\bx_{23}| e^{- i \phi_{\bx_{23}}} - z_1 |\bx_{12}| e^{- i \phi_{\bx_{12}}} )}{|\bx_{13}| e^{- i \phi_{\bx_{13}}}} 
\notag \\
a^{(+)}_{3,--} & = 
\frac{ z_1 (z_3 |\bx_{23}| e^{- i \phi_{\bx_{23}}} - z_1 |\bx_{12}| e^{- i \phi_{\bx_{12}}} )}{|\bx_{13}| e^{ i \phi_{\bx_{13}}}}, 
 \label{eq:a1L}
\end{align}
where  $\theta_p, \theta_q, \theta_k$  denote the azimuthal angle of final state momenta and $|{\bm p}|, |{\bm q}|, |{\bm k}| $ their transverse momenta.

\begin{figure*}[t]
  \centering
  \includegraphics[height=5cm]{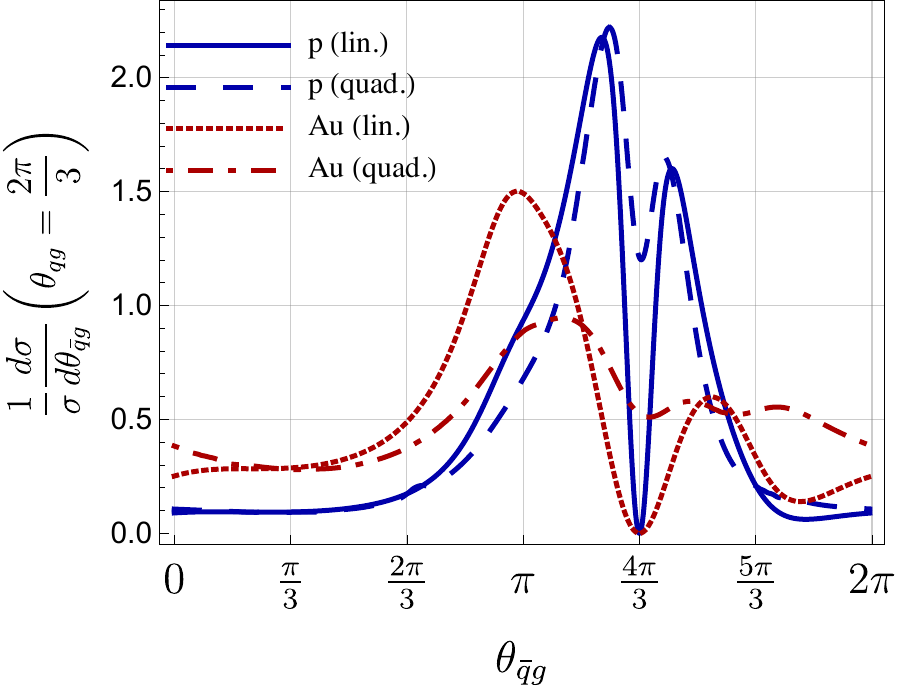} \,\,
 \includegraphics[height=5.2cm]{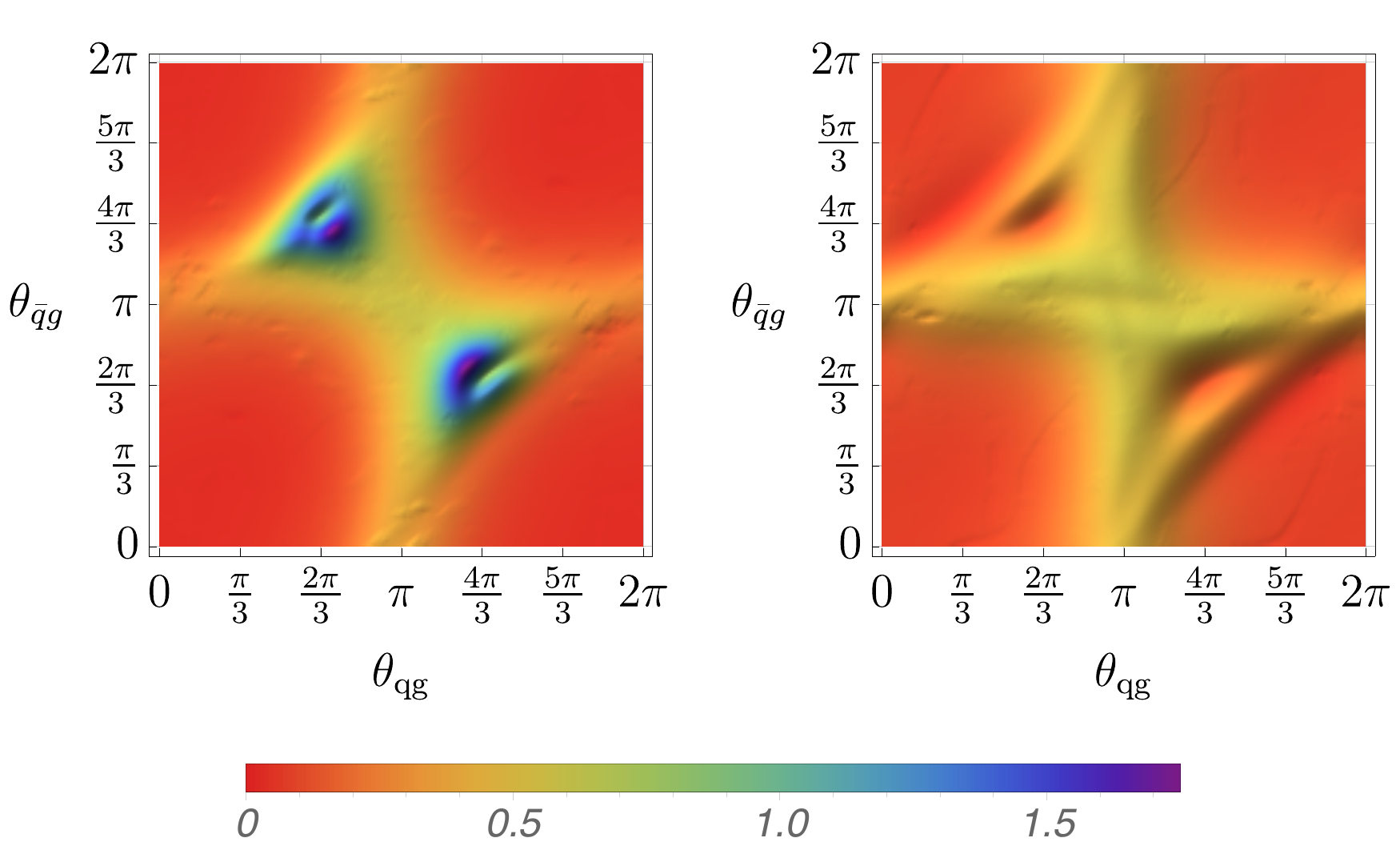}
 \caption{ We fix $z_1 = z_2 = 0.2$,
   $|{\bm p}| = |{\bm k}| = |{\bm q}| = 2$ GeV and $Q=3$ GeV. {\it
     Left:} Normalized cross-section against
   $\Delta \theta_{\bar{q}g}$ with $\Delta_{qg} = 2 \pi/3$ for proton
   and gold up to linear and quadratic order in $N^{(2)}$.  {\it
     Right:} Combined $\Delta\theta_{qg}$ and
   $\Delta\theta_{\bar{q}g}$ dependence of the normalized
   cross-section for proton and gold at quadratic order.}
  \label{fig:num1}
\end{figure*}

In absence of experimenally constrained quadrupole distributions we
use for this first study the large $N_c$ and Gaussian approximation to
write the quadrupole $S^{(4)}$ in terms of the dipole
$S^{(2)}$~\cite{cgc-double}. Furthermore, we use a model of the dipole
profile which is motivated by a fit to the solution of rcBK equation
\cite{cgc-dis}
\begin{align}
  \label{eq:20}
  S^{(2)}_{(\bx_1 \bx_2)}&= 
\int\! d^2 {\bm l} \,  e^{-i {\bm l}\cdot \bx_{12}} \, \, \Phi({\bm l}^2)
= 2  \left(\frac{ Q_0 |\bx_{12}|}{2}\right)^{\rho -1} \frac{K_{\rho -1}(Q_0  |\bx_{12}|)}{\Gamma (\rho
   -1)}, \notag \\
 \text{where} & \qquad  \Phi({\bm l}^2)  =  \frac{\rho-1}{Q_0^2 \pi } \left( \frac{Q_0^2}{Q_0^2 + {\bm l}^2}\right)^\rho\, , 
\end{align}
and $Q_0$ is a scale proportional to the saturation scale. Since we
are working in the dilute limit we study the cross-section at large
photon virtuality $Q^2=9$ GeV$^2$ and expand Eq.~\eqref{eq:19} up to
quadratic order in $N^{(2)} = 1-S^{(2)}$. The free parameters are
taken as $\rho=2.3$ and $Q_0^{\text{proton}} =0.69$ GeV which are
motivated by inclusive DIS fits of the dipole distribution at
$x=0.2 \times 10^{-3}$. For the gold nucleus we use
$ Q_0^{\text{Au}} = A^{1/6}\cdot Q_0^{\text{proton}} = 1.67$ GeV.  At
the linear order in $N^{(2)}$, the cross-section is directly
proportional to the Fourier transform of the dipole,
$\Phi\left(({\bm p} + {\bm k} + {\bm q})^2\right)$ and therefore gives
direct access to the gluon distribution in the target.  In analogy to
the back-to-back configuration in di-parton production we take
$|{\bm p}| = |{\bm k}| = |{\bm q}| $. The `collinear' limit
${\bm p} + {\bm k} + {\bm q}=0$ of vanishing transverse momentum
transfer between projectile and target corresponds then to the angular
configuration
$\{ \Delta \theta_{qg}, \Delta \theta_{\bar{q}g} \} = \{2 \pi/3, 4
\pi/3\}$
and
$\{ \Delta \theta_{qg}, \Delta \theta_{\bar{q}g} \} = \{4 \pi/3, 2
\pi/3\}$,
{\it i.e.} a Mercedes-Benz star configuration, which is characterized
by strong peaks of the angular distribution at these points. We
observe vanishing of the partonic cross section at these `collinear'
configurations, Fig.~\ref{fig:num1}, accompanied by a strong double
peak. This behavior is also observed in studies of quark-gluon,
photon-quark and dilepton-quark angular correlations \cite{q-photon}.
This vanishing of the partonic cross-section at these points is due to
the vanishing of the partonic matrix element at leading order in
$N^{(2)}$ for zero momentum transfer between projectile and
target. Indeed such a behavior is expected due to Ward identities
applicable to the gluon exchange in the $t$-channel. This double peak
will mostly go away at the hadronic level and/or when adding quadratic
corrections in $N^{(2)}$, which already provides non-zero values at
these points.  The effect of a larger gluon saturation scale for a
nucleus is clearly seen at the linear level in the figure.  To explore
the potential of the process to detect effects beyond the linear
approximation, we further include sub-leading corrections in the
dilute expansion.  We find that these corrections are small in the
case of the proton, while sizable for a highly saturated gold
nucleus. This emphasizes the potential of the 3 parton production
process in providing experimental evidence for saturation effects and
in particular for exploring the 4-point correlator $S^{(4)}$.  This is
even more remarkable due to the rather large value of photon
virtuality $Q^2 = 9$ GeV$^2$. A comprehensive numerical study of the
three hadron/jet azimuthal angular correlations using the most
up-to-date solutions of the rcBK equation will clearly help
establish/constrain saturation dynamics against competing formalism
such as collinear factorization, applicable to the low density regime.
This is work in progress and will be reported elsewhere
\cite{long:2016}.

\section*{Acknowledgments}
We would like to thank I. Balitsky, G. Beuf and Yu. Kovchegov for
helpful discussions.  We also would like to acknowldege use of the
codes \cite{codes}. Support has been received in part by Consejo
Nacional de Ciencia y Tecnolog\'ia grant number 256494 and
UNAM-DGAPA-PAPIIT grant number IN101515. M.H. acknowledges support by
Consejo Naciona de Ciencia y Tecnolog\'ia grant numbers CB-2014-22117
and CB-2014-241408 as well as the Red-FAE.  J.J-M. acknowledges
support by the DOE Office of Nuclear Physics through Grant No.\
DE-FG02-09ER41620 and from The City University of New York through the
PSC-CUNY Research Award Program, grant
67732-0045. M.E.T.-Y. acknowledges support from Consejo Naciona de
Ciencia y Tecnolog\'ia sabbatical grant number 232946 and the kind
hospitality of ICN-UNAM provided in the initial stages of this
collaboration.

\end{document}